\documentclass[a4paper,11pt]{article}
\usepackage{pos}
\usepackage{bm}
\usepackage{graphicx}
\usepackage{amsmath,mathtools} 
\usepackage{xspace}
\usepackage{wrapfig}
\usepackage{slashed}
\usepackage[normalem]{ulem}
\usepackage[utf8]{inputenc}
\usepackage{mathtools}
\usepackage{stmaryrd}
\usepackage{appendix}
\usepackage{xspace}
\usepackage{upgreek}
\usepackage{multirow}

\usepackage{scrextend}

\usepackage{tabularx}
\usepackage{tensor}

\definecolor{darkgreen}{rgb}{0.0, 0.4, 0.0}

\newcommand{\df}{\mathrm{d}}

\newcommand{\as}{\alpha_s}

\newcommand{\tb}{{\bar{t}}}

\newcommand{\GeV}{\,\mathrm{GeV}}

\def\df{\textrm{d}}

\def\mtpole{m_t^{\rm pole}}

\newcommand{\ee}{{e^+e^-}}

\allowdisplaybreaks[2]

\newcommand{\cO}{\mathcal{O}}

\usepackage[mathscr]{eucal}

\DeclareRobustCommand{\Refcite}[1]{Ref.~\cite{#1}}

\DeclareRobustCommand{\eq}[1]{eq.~\eqref{eq:#1}}

\DeclareRobustCommand{\secn}[1]{\hyperref[sec:#1]{section~\ref*{sec:#1}}}
\DeclareRobustCommand{\Sec}[1]{\hyperref[sec:#1]{Section~\ref*{sec:#1}}}

\DeclareRobustCommand{\subsec}[1]{\hyperref[subsec:#1]{subsection~\ref*{subsec:#1}}}
\DeclareRobustCommand{\Subsec}[1]{\hyperref[subsec:#1]{Subsection~\ref*{subsec:#1}}}

\DeclareRobustCommand{\app}[1]{\hyperref[app:#1]{appendix~\ref*{app:#1}}}
\DeclareRobustCommand{\App}[1]{\hyperref[app:#1]{Appendix~\ref*{app:#1}}}

\DeclareRobustCommand{\fig}[1]{\hyperref[fig:#1]{figure~\ref*{fig:#1}}}
\DeclareRobustCommand{\Fig}[1]{\hyperref[fig:#1]{Figure~\ref*{fig:#1}}}

\DeclareRobustCommand{\tab}[1]{\hyperref[tab:#1]{table~\ref*{tab:#1}}}
\DeclareRobustCommand{\Tab}[1]{\hyperref[tab:#1]{Table~\ref*{tab:#1}}}

\title{Precision Top Mass Measurement Using Energy Correlators}

\author[a,b]{Jack Holguin}
\affiliation[a]{University of Manchester,\\ School of Physics and Astronomy, Manchester, M13 9PL, United
Kingdom}
\affiliation[b]{CPHT, CNRS, \'Ecole polytechnique, Institut Polytechnique de Paris,\\ 91120 Palaiseau,
France}

\author[c]{Ian Moult}
\affiliation[c]{Department of Physics, Yale University,\\ New Haven, CT 06511}

\author*[d]{Aditya Pathak}
\affiliation[d]{Deutsches Elektronen-Synchrotron DESY,\\ Notkestr. 85, 22607 Hamburg, Germany}

\author[e]{Massimiliano Procura}
\affiliation[e]{University of Vienna, Faculty of Physics,\\ Boltzmanngasse 5, A-1090 Vienna, Austria}

\author[f]{Robert Sch\"ofbeck}
\affiliation[f]{Institute for High Energy Physics, Austrian Academy of Sciences,\\ Nikolsdorfergasse 18,
A-1050 Vienna, Austria}

\author[f]{Dennis Schwarz}

\abstract{Precision top mass measurements at hadron colliders have been notoriously difficult. The fundamental challenge in the current approaches lies in achieving simultaneously high top mass sensitivity and good theoretical control.
Inspired by the use of standard candles in cosmology, we overcome this problem by showing that a single energy correlator-based observable can be constructed that reflects the characteristic angular scales associated with both the $W$-boson and top quark.
This gives direct access to the dimensionless quantity $m_{t}/m_{W}$, from which $m_{t}$ can be extracted in a well-defined short-distance mass scheme as a function of the well-known $m_{W}$. A Monte-Carlo-based study is performed to demonstrate the properties of our observable and the statistical feasibility of its extraction from the Run 2 and 3 and High-Luminosity LHC data sets. The resulting $m_t$ has remarkably small uncertainties from hadronization effects and is insensitive to the underlying event and parton distribution functions.
Our proposed observable provides a road map for a rich program to achieve a top mass determination at the LHC with record precision.}

\FullConference{The European Physical Society Conference on High Energy Physics (EPS-HEP2023)\\
21-25 August 2023\\
Hamburg, Germany\\}

\renewcommand{\hookAfterAbstract}{\par\bigskip
\textsc{DESY-23-189}}

\renewcommand{\logo}{\relax}

\begin{document}
\maketitle

\section{Introduction}

\begin{wrapfigure}{r}{0.45\textwidth}
\vspace{-10pt}
\includegraphics[width=\linewidth]{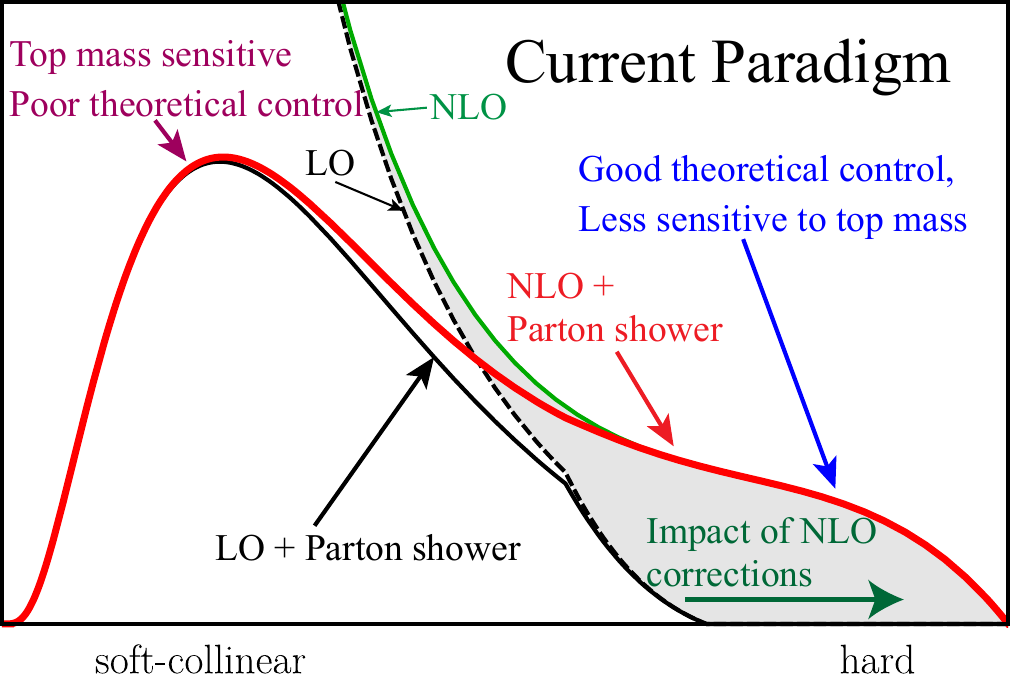}

\caption{Challenges in the current paradigm for top mass measurements. Adapted from
\Refcite{Hoang:2020iah}.}\label{fig:current}
\vspace{-4pt}
\end{wrapfigure}
Precision measurement of the top mass is crucial for the stability analysis of the electroweak vacuum, consistency tests of the SM, and indirect searches for physics beyond the SM. The top quark mass $m_t$ is not a direct physical observable but a Lagrangian parameter that receives quantum corrections that must be renormalized and defined within a specific mass scheme.
\Fig{current} represents a generic top mass sensitive observable currently employed for top mass measurement. Here, the highest mass sensitivity arises through a threshold feature such as a peaked structure or a shoulder.
The physics of the threshold region is, in general, extremely challenging to describe as it is dominated by several widely separated scales, multiple soft and collinear emissions, and large nonperturbative corrections, leading to the breakdown of fixed-order perturbation theory (as shown by the steeply rising green curve labeled `NLO' as we move towards the threshold structure). On the other hand, fixed-order corrections predominantly impact the tail where the $m_t$-sensitivity is relatively poorer.

The approach so far has been to describe the effects in the threshold region using the Monte Carlo (MC) parton shower event generators interfaced with a hadronization model. As shown in \fig{current}, the region of highest top mass sensitivity is essentially described entirely by the parton shower, and NLO corrections have little impact in this region (as shown by the closeness of the `LO+Parton Shower' and `NLO+ Parton Shower' curves in the peak region).
This has led to MC-based extractions called the \textit{direct} measurements with the smallest quoted uncertainties, with the current world average being $m_t^\mathrm{MC}= 172.76\pm 0.3$~GeV~\cite{Zyla:2020zbs}. The superscript `MC' signifies that these direct measurements are measuring the ``Monte Carlo top mass parameter'' that does not have a straightforward correspondence to a field-theoretic mass scheme definition.
While MCs are indispensable tools for LHC physics capable of describing arbitrarily complex observables, they are limited in accuracy and typically only leading color, leading logarithmic accurate.
Consequently, limitations of parton showers and their complicated interface with hadronization models have been argued to induce an additional theory/conceptual uncertainty of $\cO(1~{\GeV})$ in the top mass~\cite{Hoang:2020iah}, which is larger than the experimental uncertainty.

Next, measurements of multi-differential $t\tb$ and $t\tb + 1$ jet cross sections have yielded top mass measurements in the pole mass scheme, and the current world average stands at $\mtpole = 172.5\pm 0.7\GeV$~\cite{Zyla:2020zbs}. However, in accordance with the general features described in \fig{current}, the highest sensitivity arises in the region of low $t\tb$ masses where large threshold corrections from Coulomb and soft gluon resummation are expected to dominate. At present, they are only known with large uncertainties and are estimated to induce additional sizable $\sim 1\GeV$ uncertainty in the $\mtpole$ measurements~\cite{Piclum:2018ndt,Sirunyan:2019zvx} on top of the quoted measurement. Finally, the use of MCs also induces the top mass interpretation problem in various alternative top mass measurements that directly rely on MCs.
In conclusion, the precision measurement of the top mass with existing approaches is notoriously challenging.

\section{Energy-energy correlators on boosted top quarks}

\begin{figure}
\centering
\includegraphics[width=0.32\linewidth]{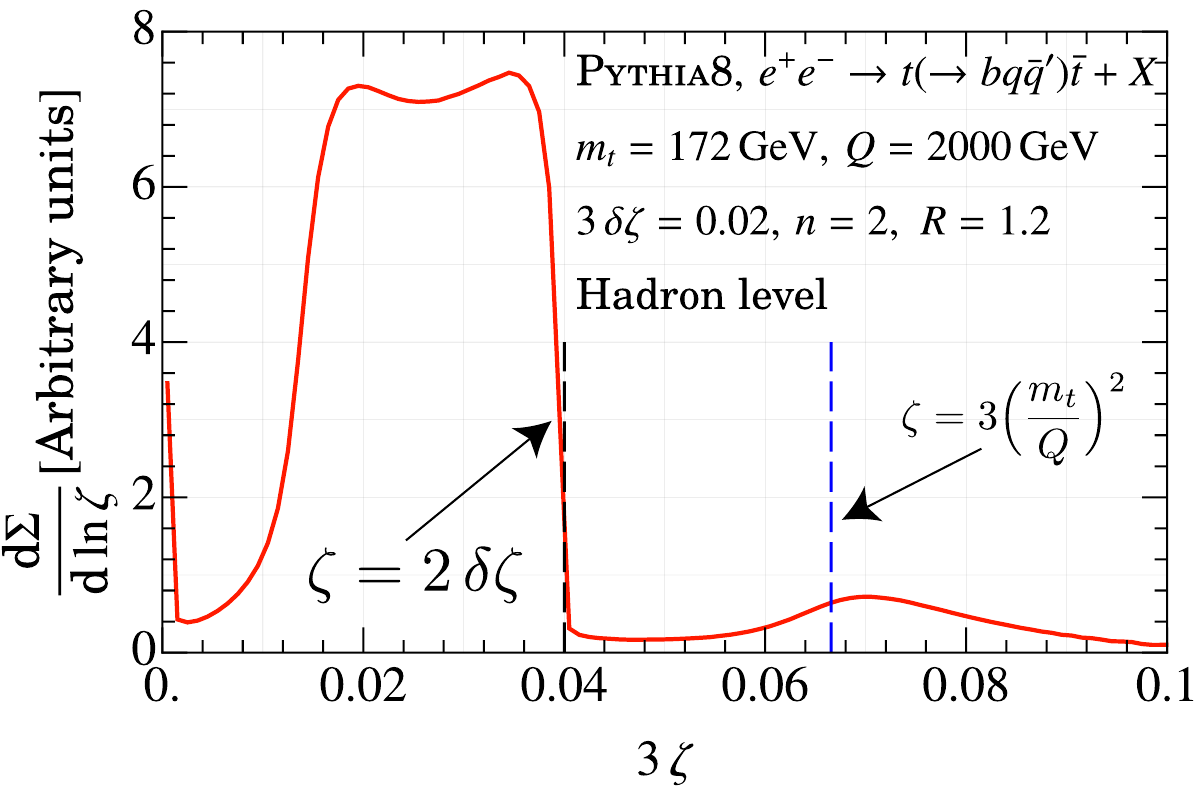}
\includegraphics[width=0.32\linewidth]{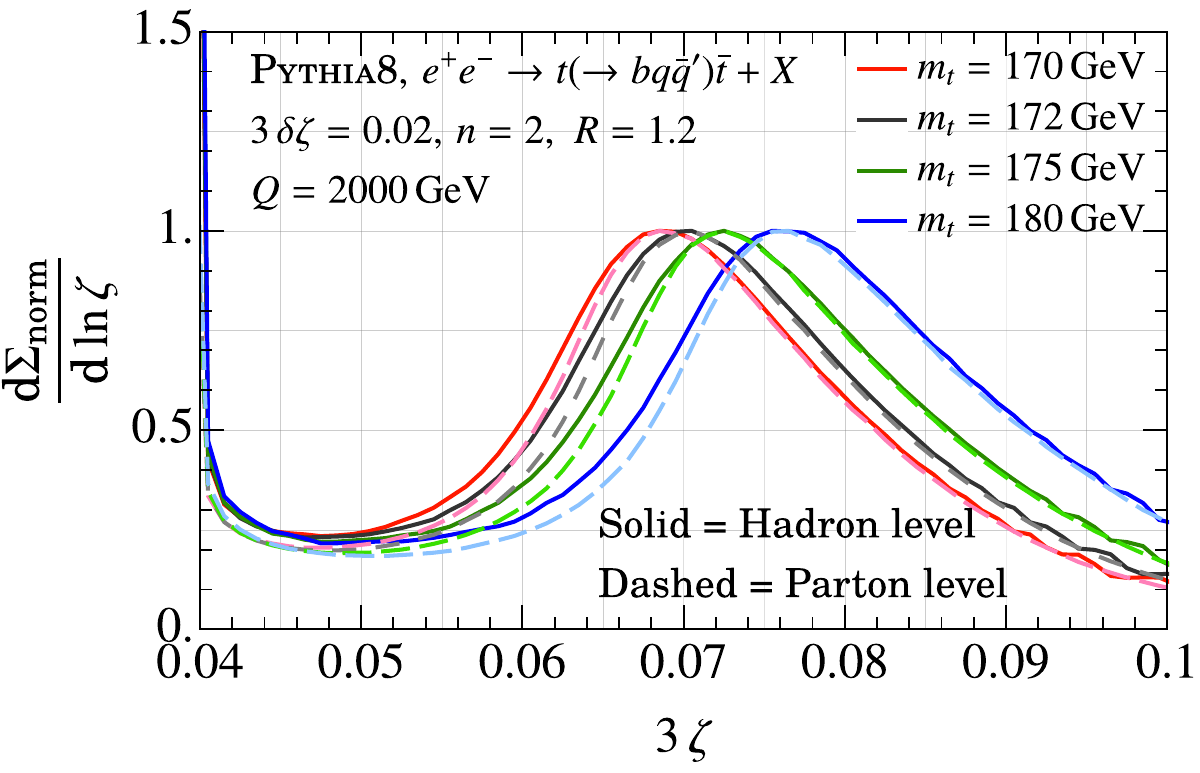}
\includegraphics[width=0.33\linewidth]{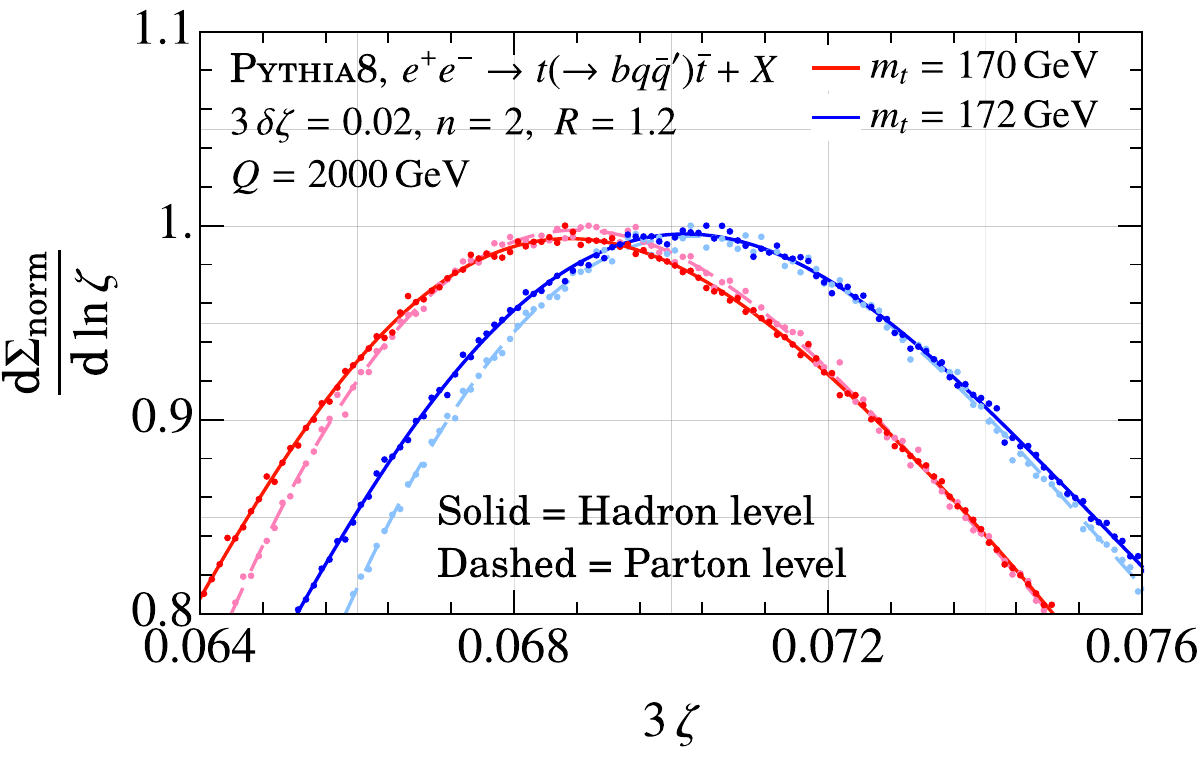}
\caption{The three-point
correlator
measured in an equilateral configuration shows the imprint of three-body decay of the top quark (left)
at
angles $\zeta_t \sim m_t^2/Q^2$ corresponding to the top quark boost. The peaked structure is
highly sensitive to the top mass (center)
and
extremely robust against hadronization (right)~\cite{Holguin:2022epo}.}\label{fig:equi}
\vspace{-15pt}
\end{figure}

Correlation functions are the most fundamental tools for specifying the ensemble properties of probabilistic systems and have a long history in QCD collider physics. Energy-Energy Correlators (EECs) quantify the angular correlations of hadronic energy flux within jets.
While EECs were already proposed several decades ago~\cite{Basham:1978bw,Basham:1977iq,Basham:1979gh,Basham:1978zq}, their unique potential for precision collider physics has only just started to emerge~\cite{Chen:2020vvp}.
Measured on a jet, EECs are given by inclusive cross sections where only the relative angles are fixed, and the energies are integrated over:
\begin{align}\label{eq:enc}
\text{E}^N\text{C} (\zeta_{12}, \zeta_{23}, \ldots )\equiv \sum_{i,j,k \in J }\int \df \zeta_{ijk\ldots} \delta
\big(\zeta_{12}- \zeta_{ij} \big) \delta(\zeta_{23} - \zeta_{kl} ) \ldots
\frac{E_i E_j E_k \ldots}{Q^N} \frac{\df \sigma_{i,j,k,\ldots} }{\df \zeta_{ijk\ldots }} \, ,
\end{align}
where E$^N$C refers to the $N$-point correlation function, and $Q$ is the large momentum (such as jet $p_T$) associated with the jet and $E_i$'s (or $p_{T,i}$'s) the energy of hadron $i$.
As \eq{enc} shows, they are inherently distinct from the typical observables mentioned above as they are not sampled event-by-event.
Just as the moments of the cosmic microwave background power spectrum offer insights into the universe's structure at various length scales in the case of cosmology, EECs map out the structure of the underlying field theory across the “sky” of the detector.
EECs are especially robust for LHC experiments: the energy weighting naturally suppresses soft extraneous contamination without requiring jet grooming. Unlike traditional jet substructure observables, they can be used directly on detector tracks with excellent angular precision without posing any limitations for theoretical calculations.
Crucially, EECs retain strong links with QFT as they can be directly mapped to correlation functions, and, therefore, a lot of fruitful connections between EECs and highly developed theoretical techniques in conformal field theories and light ray operator expansion have just begun to be explored. On the other hand, due to complicated clustering and grooming algorithms in the typical event-by-event observables, a direct link with QFT correlation functions is obscured.

In \Refcite{Holguin:2022epo}, EECs measured on boosted top quarks were proposed as a promising mass-sensitive probe.
The unique three-body (hadronic) decay of the top quark imprints itself at a characteristic angle $\zeta_t \sim m_t^2/Q^2$ in the three-point correlator (EEEC) spectrum. Specifically, in the equilateral configuration, the top quark feature in the spectrum can be isolated by imposing an asymmetry cut. This is shown in \fig{equi}, left for $\ee$ collisions. Zooming in, we see that it is highly sensitive to the top quark mass (\fig{equi}, center).
Crucially, this top-resonance feature appears in the region of the spectrum where the perturbation theory provides the dominant description. This is because, unlike the jet mass, the soft radiation enters the correlator only via recoil on the energetic hadrons. Consequently, EECs measured on the top quark are not sensitive to scales below $\Gamma_t$. This makes them naturally robust against hadronization and the underlying event (UE).
In contrast, the other jet substructure-based proposal of employing the jet mass~\cite{Fleming:2007xt,Fleming:2007qr,Bachu:2020nqn} as a $m_t$-sensitive probe, including with application of jet grooming~\cite{Hoang:2017kmk}, is sensitive to \textit{inter-jet} soft radiation at scales parametrically lower than the top width, resulting in significantly $\cO(1\GeV)$ large hadronization effects.
The right panel of \fig{equi} shows that the shift due to hadronization in the top mass is $\Delta m_t^{\text{Had.}} \sim 250$ MeV.
Hence, this idea overcomes the problems shown in \fig{current}, achieving high top mass sensitivity in the perturbative region for the first time. However, a direct application of this approach to the LHC poses challenges. Unlike the jet mass and the transverse momentum $p_T$, EECs are sensitive to a \textit{dimensionless angular scale}.
At the LHC, $Q$ must be replaced by jet $p_T$. Despite the theoretical elegance of this approach, as mentioned above, the jet $p_T$ has large experimental uncertainties and sensitivity to PDFs, making a direct extraction of the $m_t$ from a measurement of the $\zeta_t$ angle challenging.
In conclusion, identifying a top mass-sensitive observable that is simultaneously experimentally feasible at the LHC, completely robust to hadronization and the UE, and calculable to high perturbative orders has so far remained an open problem.

\begin{figure}[t]
\centering
\includegraphics[width=0.45\linewidth]{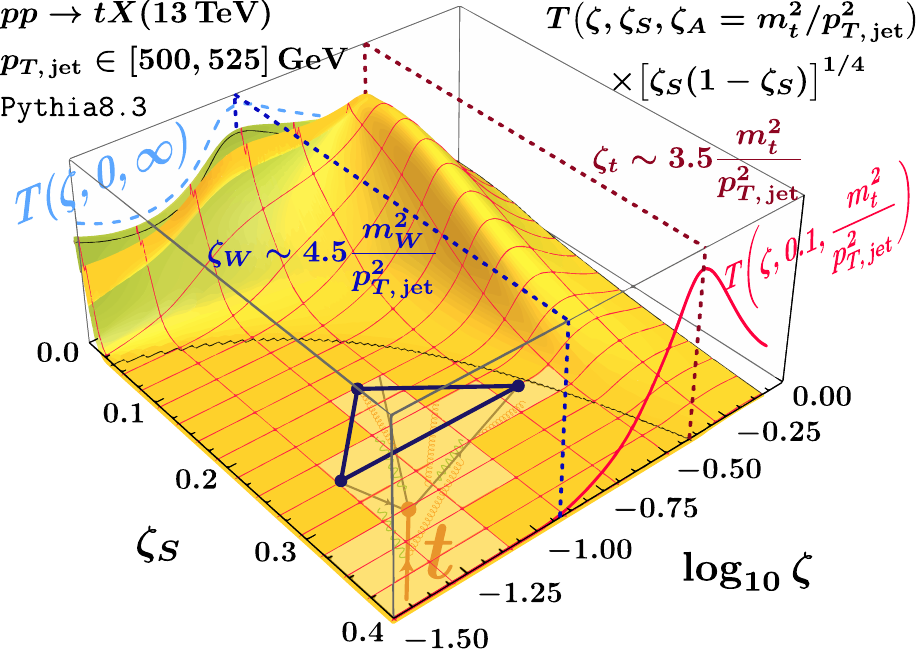}
\includegraphics[width=0.48\linewidth]{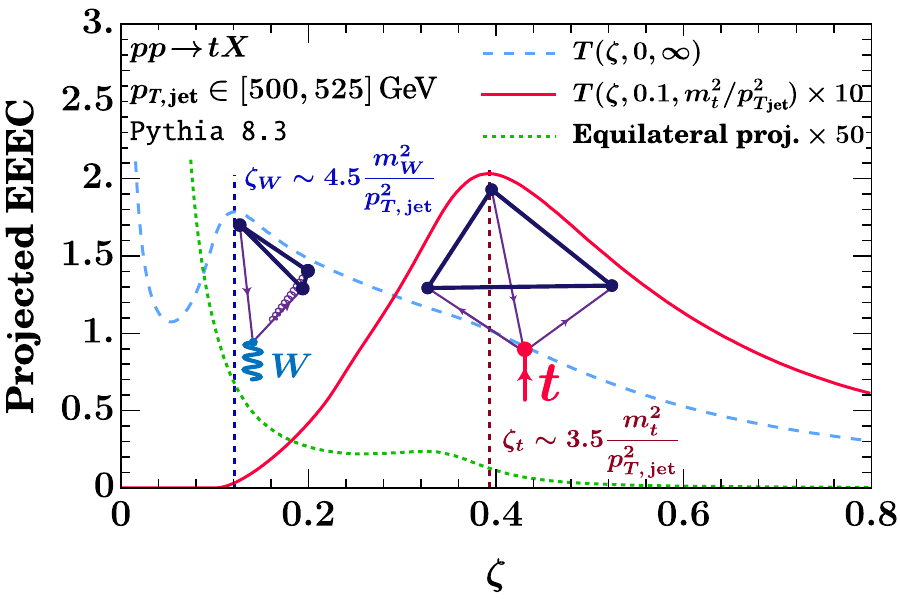}
\caption{The shape of the three-point correlator on boosted top quark jets. By lowering the small side
cut $\zeta_S$, the $W$ peak emerges~\cite{Holguin:2023bjf}. On the right, the
projections
on the front and back faces are shown, as well as the huge statistical gain compared to
the
previous equilateral projection of \Refcite{Holguin:2022epo} shown in green.}\label{fig:3d}
\vspace{-15pt}
\end{figure}

\section{The Standard Candle Approach}

In cosmology, dimensionful quantities of interest, such as distances, are not directly obtained but indirectly through precisely measured quantities, such as luminosity and standard candles. In our case, the dimensionless quantities of interest are the angles that can be measured extremely precisely, thanks to the excellent tracking system of the LHC.
Recently, \Refcite{Holguin:2023bjf} showed that a single EEC-based observable can be constructed that contains imprints from angular scales of the top quark and $W$ boson decays. It is a specific integrated EEC cross-section $T(\zeta, \zeta_S, \zeta_A)$ where the angle $\zeta$ is an average of the medium and long sides of the triangle on the celestial sphere probed by the correlator, and its shape is shown in \fig{3d}.
Here, the medium and long sides are required not to differ more than the asymmetry cut $\zeta_A$, and the short side is required to be larger than the $\zeta_S$. The plot of this distribution shown in \fig{3d} reveals an interesting shape: for large $\zeta_S$, there is a single peak at the angular scale associated with the top decay $\zeta_t \sim m_t^2/p_T^2$. Upon lowering $\zeta_S$, while keeping $\zeta_A$ fixed, we notice an emergence of a peak at the angular scale of the $W$, $\zeta_W \sim m_W^2/p_{T}^2$.
This results from the imprint of two-body resonant decay of the $W$ boson in the ``squeezed configuration'' when one of the angles is much smaller than the other two.
In the right plot in \fig{3d}, we show the projections on the front and back faces.

\begin{wrapfigure}{r}{0.48\textwidth}
\centering
\includegraphics[width=\linewidth]{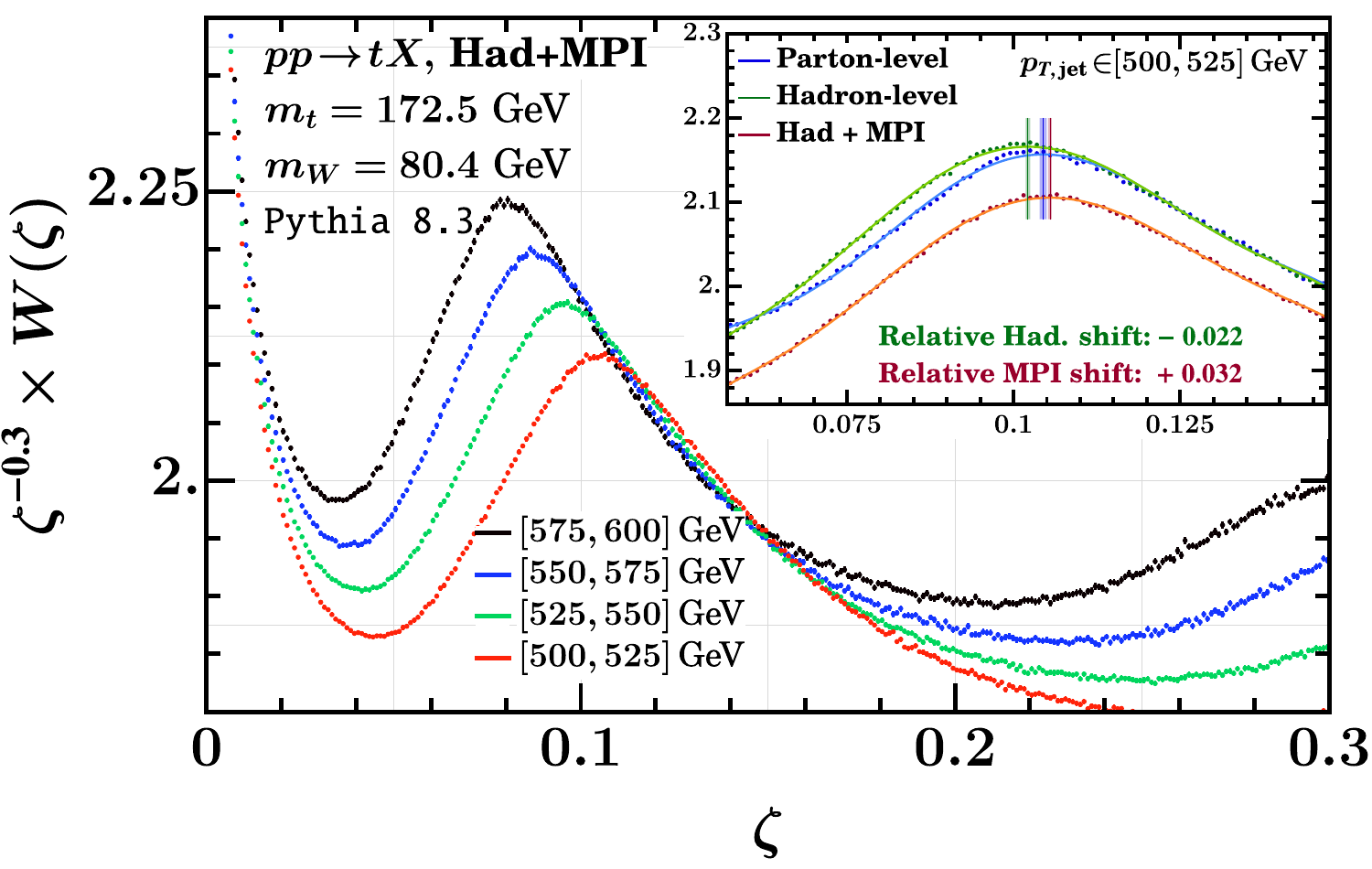}
\vspace{-20pt}
\caption{The ratio of three-point to two-point EEC provides a robust standard candle for identifying the
angular scale associated with the $W$.}\label{fig:wratio}
\vspace{-9pt}
\end{wrapfigure}
The presence of the $W$ in the distribution introduces dependence on the $m_t/m_W$ ratio, which can
be extracted from the measurements around the two peaks in the three-point EEC.
This shows that despite the complexity of the three-body decay of the top quark, the intermediate $W$ boson is, in fact, an important gift, enabling an elegant and complete calibration mechanism.
The standard candle observable robust to hadronization and underlying event effects and calculable to high perturbative orders is shown in \fig{wratio}. Here, instead of directly using the projection shown in \fig{3d}, we consider a ratio of the three-point to two-point EEC measured on top quark jets (removing any restriction of the short side and asymmetry cuts), $W(\zeta)$. This is because, unlike for the top, where $\Gamma_t$ provides a natural cutoff, the $W$ boson decay is exposed to nonperturbative scales. The dominant nonperturbative effects drop out when considering such a ratio $W(\zeta)$. In light quark and gluon jets, these ratios have been computed to high perturbative orders and led to the most precise jet substructure $\as$ determination at the LHC~\cite{CMS:2023wcp}. This is the first time such a standard candle approach has been used in the precision jet substructure measurements. This is entirely distinct from the direct $m_t$ measurements, where the $W$ decay is reconstructed solely to achieve a fine-grained calibration of energy scales to reduce experimental uncertainties.

\section{Robustness against hadronization, UE, and initial state effects}

\begin{figure}[t]
\centering
\includegraphics[width=0.45\linewidth]{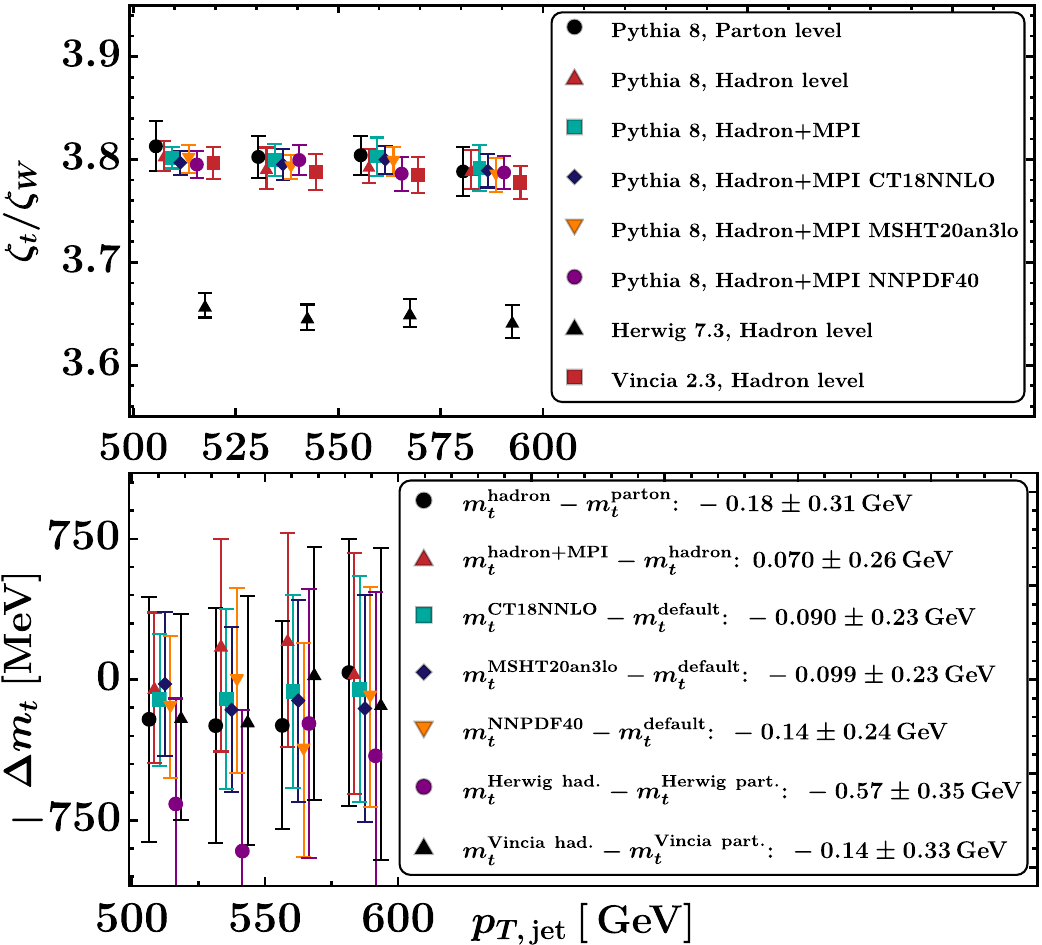}
\includegraphics[width=0.54\linewidth]{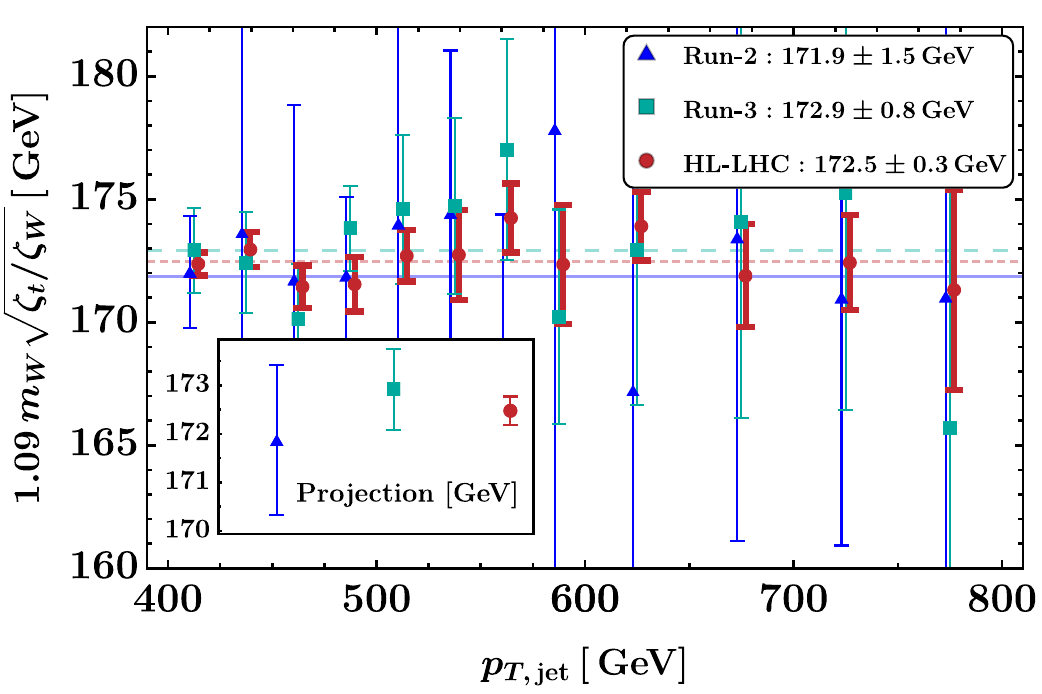}
\caption{Simulations showing remarkable properties of the ratio $\zeta_t/\zeta_W$ in terms of
independence of the $p_{T,\, \text{jet}}$ (left, top), robustness against hadronization, the UE and
variations in PDF sets (left, bottom), and promising prospects of a top mass extraction at the
HL-LHC
with
record precision (right).}\label{fig:eecpT}
\end{figure}

The $W$-resonance feature captured through the standard candle observable $W(\zeta)$ is highly correlated with the top quark, such that in the ratio $\zeta_t/\zeta_W$, the dependence on the $p_T$, as well as the systematic effects that impact $p_T$, entirely cancels.
This is because a single shared $p_T$-dependent boost determines the peak locations. This was thoroughly verified in \Refcite{Holguin:2023bjf} through extensive simulation studies. The results are shown in \fig{eecpT}.
The top left panel shows that the ratio is essentially independent of the $p_T$.
Perturbative corrections modify the ratio's precise value and are directly sensitive to the ratio $m_t/m_W$.

Crucially, while the jet $p_T$ itself is significantly impacted by hadronization and the UE and sensitively depends on the choice of PDFs, these effects cancel to a remarkable degree in the ratio. We see this in the left panel, bottom. Here, the relative shifts in the peak locations are represented as equivalent shifts in the extracted top mass $\Delta m_t$.
Variations in PDFs impact the $p_T$ spectrum, leading to shifts equivalent to a few GeVs in each peak individually. Still, in the combination, we see a remarkable cancellation down to almost 200 MeV (consistent with zero shift within the error bars).
Likewise, the impact of the UE, which can be as large as $3\%$ in the individual $W$ and top peaks (also shown in \fig{wratio} in the inset for the $W$ peak), or equivalently $2.5\GeV$ shift in the top mass, is completely eliminated in this standard candle approach (the second entry in the middle panel with $\Delta m_t = 70 \pm 260$ MeV). Finally, we also see a robust behavior against hadronization. While the hadronization shifts here can be somewhat larger, $\sim 500$ MeV, the approach developed in the second part of this proposal, will yield powerful methods to account for this adequately.
Nevertheless, they are still significantly smaller than $\sim 1\GeV$ nonperturbative shifts in the peak of the groomed jet mass distribution.
Finally, this observable must be precisely measurable with the data collected during HL-LHC. The right panel of \fig{eecpT} shows an estimate of statistical uncertainty at
Runs 2 and 3 of the LHC and the HL-LHC demonstrate this is possible. These results forecast statistical precision on $m_{t}$ at a few hundred MeV level for the HL-LHC and with $1 \GeV$ or better precision with the LHC luminosities.

In conclusion, the promising results in the above simulation studies provide a strong motivation for investing effort into developing this standard candle approach.
This will necessitate the development of a new formalism to describe exclusive top decay processes.
In the simulations discussed above, the ratio $\zeta_t/\zeta_W$ was accessed by fitting the peak
locations. A theory prediction for the full shape near the top and $W$ resonance peaks, on the other
hand, will enable a
much finer control over the observable, resulting in an increased precision on the extracted $m_t$ and
a consistent treatment of perturbative uncertainties. While this will be challenging, the required effort will
be
well worth the reward.

\acknowledgments
I.M. is supported by start-up funds from Yale University. J.H. is supported
by the Leverhulme Trust as an Early Career Fellow. This work is supported in part by the GLUODYNAMICS
project funded by the ``P2IO LabEx (ANR-10-LABX-0038)'' in the framework ``Investissements d'Avenir''
(ANR-11-IDEX-0003-01) managed by the Agence Nationale de la Recherche (ANR), France.

\bibliographystyle{JHEP}
\bibliography{top3}

\end{document}